\documentclass[runningheads,a4paper]{llncs}

\usepackage{amssymb}
\usepackage[export]{adjustbox}
\usepackage{amsmath}
\usepackage{graphicx}
\usepackage[multi-part-units=single]{siunitx}
\DeclareSIUnit\pixel{pixel}
\usepackage{bm}
\usepackage{amsbsy}
\usepackage{url}
\usepackage{multirow}
\usepackage{tikz}

\newcommand{\keywords}[1]{\par\addvspace\baselineskip
\noindent\keywordname\enspace\ignorespaces#1}

\begin{document}

\mainmatter  % start of an individual contribution

\title{Left Ventricle Quantification Using Direct Regression with Segmentation Regularization and Ensembles of Pretrained 2D and 3D CNNs}

\titlerunning{Left Ventricle Quantification with Pretrained CNNs}


\author{Nils Gessert\inst{1} \and Alexander Schlaefer\inst{1}}
%\author{Anonymous Submission}

\authorrunning{Nils Gessert et al.} 
%\authorrunning{}



\institute{$^1$Institute of Medical Technology, Hamburg University of Technology, Hamburg, Germany\\
	\email{nils.gessert@tuhh.de}\\
}
%\institute{}


\maketitle

\begin{abstract} %TODO at least 150 and at most 250 words
%The abstract should summarize the contents of the paper and should
%contain at least 70 and at most 150 words. It should be written using the
%\emph{abstract} environment.

%Cardiac left ventricle (LV) quantification can be useful for diagnosing cardiac diseases. Automatic calculation of all relevant LV indices from cardiac MRI sequences is a challenging task due to large variation between patients and deformation during the cardiac cycle. Previous methods have typically approached LV quantification by direct regression from sequences of MRI images, by segmenting the myocardium first or a combination of both. To handle temporal relationships, either recurrent models or spatio-temporal convolutional neural networks (CNNs) have been employed. In the context of the LVQuan19 Challenge, we present a new approach inspired by previous methods. We put a strong emphasis on transfer learning as the training dataset is very small. We consider pretraining both for spatial 2D and spatio-temporal 3D CNNs which directly regress LV indices and classify the cardiac phase. To incorporate segmentation information, we propose an architecture-independent segmentation regularization. We improve performance further with a new ensembling strategy where we automatically select the optimal combination of models for each subtask. 

Cardiac left ventricle (LV) quantification provides a tool for diagnosing cardiac diseases. Automatic calculation of all relevant LV indices from cardiac MR images is an intricate task due to large variations among patients and deformation during the cardiac cycle. Typical methods are based on segmentation of the myocardium or direct regression from MR images. To consider cardiac motion and deformation, recurrent neural networks and spatio-temporal convolutional neural networks (CNNs) have been proposed. 
We study an approach combining state-of-the-art models and emphasizing transfer learning to account for the small dataset provided for the LVQuan19 challenge. We compare 2D spatial and 3D spatio-temporal CNNs for LV indices regression and cardiac phase classification. To incorporate segmentation information, we propose an architecture-independent segmentation-based regularization. To improve the robustness further, we employ a search scheme that identifies the optimal ensemble from a set of architecture variants. Evaluating on the LVQuan19 Challenge training dataset with 5-fold cross-validation, we achieve mean absolute errors of $\SI{111 \pm 76}{\milli\metre^2}$, $\SI{1.84 \pm 0.9}{\milli\metre}$ and $\SI{1.22 \pm 0.6}{\milli\metre}$ for area, dimension and regional wall thickness regression, respectively. The error rate for cardiac phase classification is $\SI{6.7}{\percent}$.

\keywords{Left Ventricle Quantification, Transfer Learning, Regression, Regularization}
\end{abstract}

\section{Introduction}

Left ventricle (LV) quantification from cardiac magnetic resonance imaging (MRI) data is often employed for assessment of cardiac function and for diagnosing diseases \cite{karamitsos2009role}. The relevant LV indices include the myocardium and cavity area, three LV cavity dimensions, six regional wall thickness (RWT) parameters and the cardiac phase (systole and diastole). In practice, LV indices are usually obtained by manual segmentation of the myocardium which is time-consuming and associated with a high intra- and inter-observer variability \cite{suinesiaputra2015quantification}. 

In recent years, a lot of work has gone into automatic LV indices estimation which is challenging due to high variability of cardiac structure between patients and deformation during the cardiac cycle. To overcome these problems, deep learning methods have been employed as they have shown success for a variety of image-based learning problems. One approach is to segment the myocardium with a convolutional neural network (CNN) and calculate relevant metrics afterwards \cite{avendi2016combined,ngo2017combining}. Alternatively, LV indices can be regressed directly from the images \cite{xue2017full,xue2018full,li2018left}. 
Other methods have combined segmentation and regression, e.g., by regressing indices from a segmentation with an end-to-end model \cite{wang2019quantification} or by adding a regression path to a segmentation model \cite{xu2018calculation}.

Besides incorporating segmentation and direct estimation, handling temporal dependencies is important as well. Often, temporal dependencies are modeled using recurrent neural networks which have also been employed for LV quantification \cite{xue2018full}. Another approach is to utilize spatio-temporal 3D CNNs to capture the relation between temporal slices \cite{jang2018full}.

In this work, we describe a new approach for LV quantification in the context of the LVQuan19 Challenge. In contrast to a majority of previous work, the associated dataset is significantly smaller (56 patients) and the MRI images are hardly preprocessed with a high spatial resolution but without any region of interest (ROI) cropping. Thus, an algorithm needs to deal with small dataset size and make use of the high image resolution while focusing on the relevant region in the image.

To address these challenges, we take previous approaches into account while putting a strong focus on using pretrained CNNs. Using transfer learning from ImageNet has been successful for a lot of medical imaging modalities, particularly when data is as scarce as in our case \cite{shin2016deep,gessert2019.ivoct}. Models trained for the common problem of image classification can be adapted for regression by replacing the output layer. Thus, we perform direct LV indices regression using various pretrained CNNs. For temporal processing, we employ spatio-temporal 3D CNNs. We enable effective training of these parameter-intensive 3D CNNs with an initialization strategy where we assign pretrained 2D weights to 3D kernels. We address high spatial image resolution paired with uncertain ROIs by using a multi-crop evaluation strategy that covers the entire image. To incorporate segmentation information into predefined models, we propose an architecture-independent regularization by adding a decoder to the model. Finally, we integrate all our models with a new ensembling approach where we automatically select the best performing ensemble for each regression and classification task.

\section{Methods}

\subsection{Dataset and Preprocessing}

The LVQuan19 training dataset consists of short-axis MRI data from $56$ patients. For each patient $20$ slices representing one cardiac cycle are provided. The resolution of the MRI slices is either $256\times 256$ or $512\times 512$ with a pixel spacing between $\SI[per-mode=fraction]{0.6836}{\milli\metre\per\pixel}$ and $\SI[per-mode=fraction]{1.7188}{\milli\metre\per\pixel}$. For each slice, segmentation masks of the myocardium and cavity area, all $11$ LV indices and the cardiac phase are provided. The LVQuan19 Challenge goal is the estimation of all $11$ indices and the cardiac phase, the segmentation masks can be optionally used.

First, we resize all images to have a pixel spacing of $\SI[per-mode=fraction]{1}{\milli\metre\per\pixel}$. Second, we take a center crop of size $300\times 300$ which is the smallest image size of all resized slices. We clip intensities between the 1st and 99th percentile. Afterwards, we perform image normalization by subtracting the mean and dividing by the standard deviation for each slice. Then, the intensities are scaled between $0$ and $1$. We convert the target indices to $\si{\milli\metre}$. Then, we scale all regression targets to a range of $0$-$1$ for training. For evaluation, the targets are scaled back to their original range. We split the dataset into $5$ cross-validation (CV) folds.

\subsection{Models} \label{sec:methods}

\begin{figure}[ht]
\begin{center}
   \includegraphics[width=1.0\linewidth]{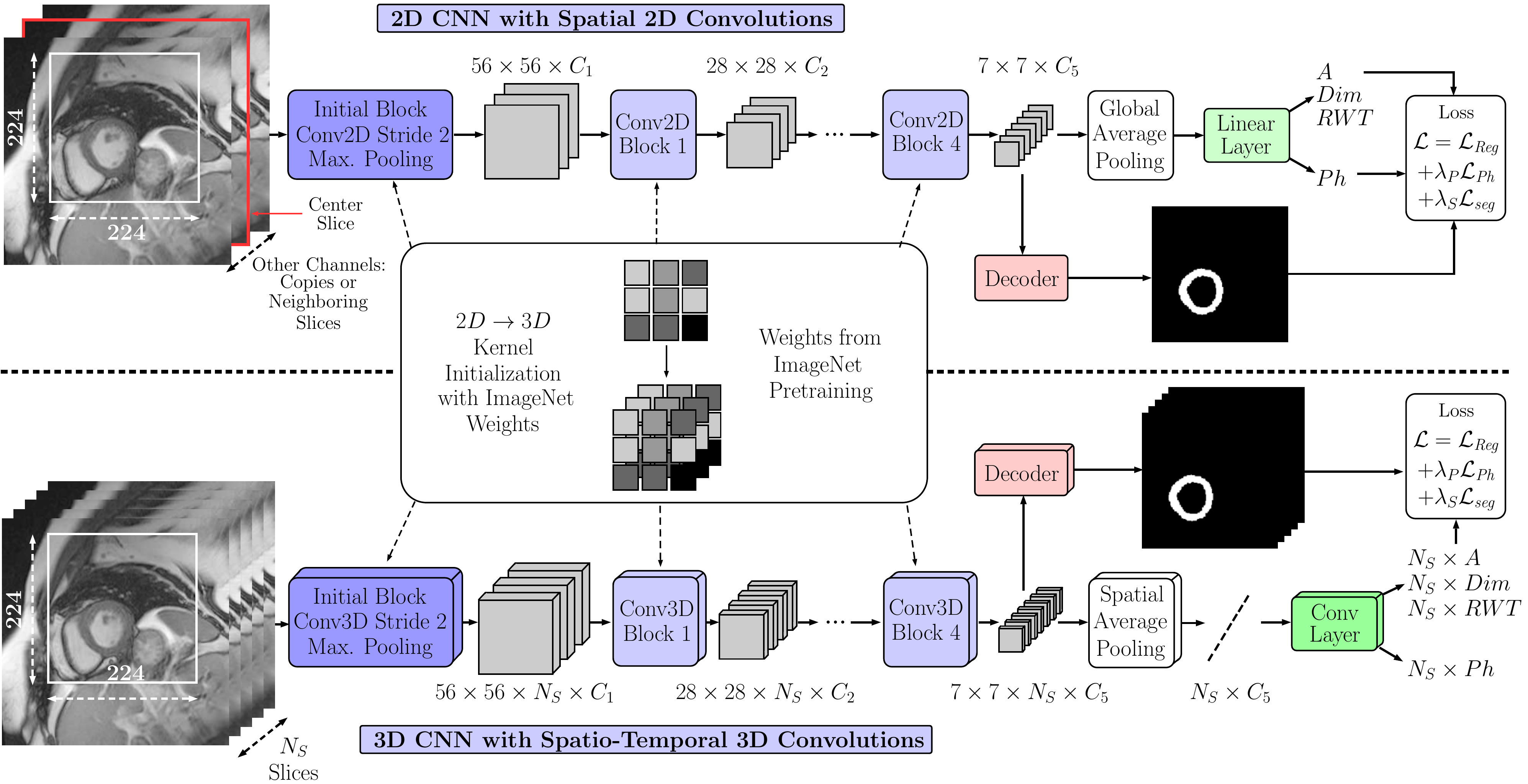}
\end{center}
   \caption{Overview of our approach. We use both 2D CNNs with 2D slices (top) and 3D CNNs with temporally stacked slices (bottom). Both are initialized with pretrained weights from ImageNet. The initial and Conv2D/Conv3D blocks have a different structure based on the respective architecture.} 
\label{fig:approach}
\end{figure}

\textbf{2D CNN Approaches.} The key idea of our approach is to use pretrained architectures for indices regression. Thus, we consider a pool of pretrained architectures including Densenet (DN) \cite{huang2017densely}, Resnet (RN) \cite{he2016deep}, Resnext (RX) \cite{xie2017aggregated} and Squeeze-and-Excitation Networks (SE) \cite{hu2018squeeze}. The overall approach is shown in Figure~\ref{fig:approach}. 

Each CNN was pretrained on ImageNet for classification of natural images into $1000$ classes. We replace the model's output layer to match the number of outputs for our problem. We consider regression only with $11$ outputs, classification (cardiac phase) with $2$ softmaxed outputs or both simultaneously with $13$ outputs. The pretrained model expects a 3-channel input image which we handle with two approaches. Either, we use a single slice, copied to all three channels or we include the two neighboring slices.

\textbf{3D CNN Approaches.} We also consider spatio-temporal 3D CNN approaches. We hypothesize that temporal context might improve indices regression. Furthermore, cardiac phase classification requires temporal context and including more slices might be helpful. The 3D CNN input is of size $H\times W\times N_S$ where $H$ and $W$ are the spatial slice dimensions and $N_S \in [1,20]$ is the number of selected slices. We also employ pretrained models to tackle small dataset size. Thus, we reuse the same pool of 2D CNNs and extend them to 3D by replacing all 2D operations by their 3D counterpart. The 3D convolutional kernels are initialized by copying the 2D kernels pretrained on ImageNet of size $h_c\times w_c$ several times into the new kernel of size $h_c\times w_c \times d_c$. To ensure consistent value ranges, the new kernels' weights are multiplied by $1/d_c$. Throughout the entire network, we do not change the slice dimension, i.e., we produce $N_S$ predictions for $N_S$ input slices in a single forward pass. For this purpose, we replace the linear output layer by a convolutional layer with kernel size $1$ which is able to handle arbitrarily-sized inputs. Due to increase in memory requirements, we only consider 3D variants of the smaller CNNs in our pool which includes Densenet121, Densenet161, and SE-Resnet101. 

\textbf{Segmentation Regularization}. In addition, we propose a segmentation-based regularization (SR). Here, we add an additional decoder to the architecture, before its global average pooling (GAP) layer. The decoder upsamples the spatial dimension of the feature maps in several steps until the original input image size is reached. At each step, we apply a convolution with kernel size $1$ which halves the current feature map dimension, followed by nearest neighbor upsampling with a factor of $2$. Then, a standard Resnet block is applied. In total, there are $5$ upsampling stages. At the output, we predict a softmaxed probability map which is used to calculate a cross-entropy loss with the ground-truth masks. During training, the loss is propagated through the entire network, forcing the core architecture to learn features for segmentation and indices regression simultaneously. We do not use the predicted segmentation explicitly for indices calculation, it only serves as a regularizer for the network. We employ this regularization strategy both for 2D and 3D CNNs.

\textbf{Data Augmentation.} Due to the small dataset size, we employ extensive data augmentation. We use random rotation with $\theta \in [\SI{0}{\degree},\SI{360}{\degree}]$ which we found more effective than simple $\SI{90}{\degree}$ rotations. Furthermore, we employ random scaling by resizing the slices by a factor of $s_c \in [0.8,1.2]$ with appropriate cropping and zero padding, if required. The targets are scaled accordingly (quadratic for areas). We chose a small batch size of $b=8$ to induce more variation during training. We did not see any improvement for dropout, $L1$ or $L2$ regularization. 

\textbf{Model Input Strategy.} The pretrained models' standard input size is $224\times 224$ while our preprocessed images are of size $300\times 300$. Therefore, we crop patches from random image locations during training. This should have a regularizing effect as the CNN gets more robust towards different relative LV locations in the images. For 2D CNNs, we randomly select a cropped slice from $b$ different patients to construct a batch of size $b$. For 3D CNNs, we randomly select a sequence of $N_S$ slices from $b$ patients for each batch. For evaluation, we follow a multi-crop evaluation approach to cover the entire image \cite{gessert.2019skin}. We crop from $4$ predefined locations in each slice and average the result. For 2D CNNs, this is repeated for every slice for each patient. For 3D CNNs, we crop sequences of $N_S$ slices from every possible location $n_s \in [1,20]$. Then, we average over all overlapping regions to obtain a final prediction for each slice. To handle the start and end of the sequence, we use cyclic repetition.

\textbf{Training.} In total, our loss function consists of the mean squared error (MSE) for regression, the cross-entropy (CE) loss for phase classification and the CE loss for segmentation. The two CE losses are not present in every model. If they are included, phase classification is weighted by $\lambda_P = 0.05$ and the segmentation loss is weighted by $\lambda_S = 0.1$. The final loss is the sum of all individual loss functions. For optimization we employ Adam with a learning rate of $l_r = \num{1e-4}$. We train each CV model for $150$ epochs where a single epoch consists of $10$ random crops from each patient in the training set. Overall, we train multiple models with different configurations $s_i$. The configuration options include the network architecture (e.g. Densenet121), the input dimension (2D or 3D), $N_S \in \{3,5,7,10\}$ for 3D CNNs, weight initialization (random or ImageNet), segmentation-based regularization (on or off) and prediction targets (areas, dimensions, RWTs, phase).

\textbf{Ensembling.} Instead of deciding for a single model we search for the optimal ensemble. Let $S=\{s_1, \dots, s_n\}$ be the set of all configurations we consider, and let $V=\{v_1, \dots, v_m\}$ be the set of all CV splits. Then we obtain predictions $\hat{y}_{ij} = f(s_i,v_j)$ for all combinations of $i$ and $j$ after training configuration $s_i$ on $V \setminus {v_j}$. We obtain the predictions per configuration through concatenation as $\hat{y}_i = \cup_j \hat{y}_{ij}$. Subsequently we perform an exhaustive search to identify the set $S* \subseteq S$ such that $\hat{y}* = \frac{1}{|S*|}\sum_{k \in S*}\hat{y}_k$ minimizes the error between $\hat{y}*$ and ground-truth $y$. For the challenge test set we obtain predictions with all models in the optimal subset $S*$ which are subsequently averaged into a final prediction. We search for the optimal subset for each task (areas, dimensions, RWTs, phase) individually to maximize performance. To keep the search time bounded, $S$ only includes our top $20$ configurations, ranked by individual CV performance.

Our proposed strategy potentially leads to implicit overfitting of the subset choice to the CV sets. Therefore, results reported for the CV sets might overestimate the performance gain of our ensembling strategy. We overcome this problem by introducing an additional test split for ensemble evaluation only. Here, we split each CV fold into two folds. We use the first portion of these new sets to find the optimal subset with our strategy. Then, we use the second portion of sets to evaluate the strategy. 

%To merge all our trained models, we propose a new ensembling technique. Each training with a configuration results in $5$ models for $5$ CV splits. For each configuration we obtain predictions from its $5$ CV models and concatenate the predictions. This results in a single prediction vector for each configuration. To find the optimal subset of configurations, we consider all possible combinations between them. For every combination, we average the predictions over the respective subset and calculate the errors with the averaged predictions. Then, we select the optimal combination based on the lowest error. The final predictions on the unseen test data are then performed with the best performing subset of configurations. We search for the optimal subset for each task (areas, dimensions, RWTs, phase) individually to maximize performance.

\begin{table}[t]
\setlength{\tabcolsep}{3.5pt}
\caption{All results for different configurations. We consider the mean absolute error (MAE) with standard deviation in $\si{\milli\metre}$ ($\si{\milli\metre}^2$ for areas) and Pearson correlation coefficient (PCC) for regression and the error rate (ER) for classification. Configurations include no pretraining (nopre), segmentation regularization (SR), joint indices regression and phase classification (Joint) and phase classification only (Class.). For ensembling, we consider taking the average over all models (Average) and our new strategy (Optimal). Results marked with a star (*) are evaluated on a different test split, see Section~\ref{sec:methods}. We use models based on Densenet (DN), Resnet (RN) and Resnext (RX).}
%For phase classification with 2D models, we stack neighboring slices in the channel dimension.
\begin{center}
\begin{tabular}{l l l l l l l l}
\hline
\textbf{Configuration} & \multicolumn{2}{c}{\textbf{Areas}} & \multicolumn{2}{c}{\textbf{Dimensions}} & \multicolumn{2}{c}{\textbf{RWTs}} & \textbf{Phase}  \\
 & MAE & PCC & MAE & PCC & MAE & PCC & ER \\
\hline
 DN121 2D nopre & $199 \pm 129$ & $0.935$ & $2.98 \pm 1.8$ & $0.945$ &  $1.55 \pm 0.9$ & $0.770$ & - \\
 DN121 2D & $139 \pm 74$ & $0.972$ & $2.38 \pm 1.3$ & $0.967$ & $1.33 \pm 0.7$ & $0.835$ & - \\
 DN121 2D SR & $\pmb{133 \pm 76}$ & $\pmb{0.974}$ & $\pmb{2.08 \pm 1.2}$ & $\pmb{0.975}$ & $\pmb{1.30 \pm 0.7}$ & $\pmb{0.847}$ & - \\ 
 DN121 2D Joint & $161 \pm 89$ & $0.960$ & $2.54 \pm 1.3$ & $0.963$ & $1.39 \pm 0.7$ & $0.823$ & $9.0$ \\
 DN121 2D Class. & - & - & - & - & - & - & $\pmb{8.4}$ \\ \hline
 DN121 3D nopre & $180 \pm 165$ & $0.940$ & $2.77 \pm 2.4$ & $0.943$ & $1.48 \pm 0.8$ & $0.798$ & - \\
 DN121 3D & $133 \pm 82$ & $0.971$ & $2.26 \pm 1.3$ & $0.968$ & $1.32 \pm 0.7$ & $0.838$ & - \\ 
 DN121 3D SR & $\pmb{126 \pm 71}$ & $\pmb{0.975}$ & $\pmb{2.14 \pm 1.2}$ & $\pmb{0.972}$ & $\pmb{1.30 \pm 0.8}$ & $\pmb{0.844}$ & - \\ 
 DN121 3D Joint & $146 \pm 74$ & $0.966$ & $2.33 \pm 1.2$ & $0.969$ & $1.43 \pm 0.8$ & $0.812$ & $7.9$ \\ 
 DN121 3D Class. & - & - & - & - & - & - & $\pmb{7.5}$ \\ \hline
 DN169 2D SR & $122 \pm 72$ & $0.976$ & $1.99 \pm 1.2$ & $0.976$ & $1.27 \pm 0.7$ & $0.853$ & - \\ 
 DN161 2D SR & $127 \pm 81$ & $0.971$ & $2.00 \pm 1.1$ & $0.978$ & $1.30 \pm 0.7$ & $0.843$ & - \\
 SE-RN101 2D SR & $126 \pm 70$ & $0.974$ & $2.01 \pm 1.1$ & $0.977$ & $1.32 \pm 0.8$ & $0.844$ & - \\    
 SE-RN152 2D SR & $124 \pm 81$ & $0.974$ & $2.07 \pm 1.2$ & $0.975$ & $1.29 \pm 0.7$ & $0.849$ & - \\ 
 RX101-64d 2D SR & $\pmb{118 \pm 72}$ & $\pmb{0.976}$ & $\pmb{1.99 \pm 1.0}$ & $\pmb{0.978}$ & $\pmb{1.25 \pm 0.7}$ & $\pmb{0.859}$ & - \\
 SE-RX101 2D SR & $135 \pm 80$ & $0.971$ & $2.23 \pm 1.2$ & $0.973$ & $1.33 \pm 0.8$ & $0.839$ & - \\  
 SENet154 2D SR & $129 \pm 81$ & $0.973$ & $2.12 \pm 1.3$ & $0.973$ & $1.31 \pm 0.8$ & $0.846$ & - \\ \hline
 Ensemble Average & $118 \pm 76$ & $0.977$ & $1.96 \pm 1.1$ & $0.978$ & $1.26 \pm 0.6$ & $0.856$ & $7.7$ \\
 Ensemble Optimal & $\pmb{111 \pm 76}$ & $\pmb{0.979}$ & $\pmb{1.84 \pm 0.9}$ & $\pmb{0.980}$ & $\pmb{1.22 \pm 0.6}$ & $\pmb{0.864}$ & $\pmb{6.7}$ \\ 
  Ensemble Average* & $117 \pm 75$ & $0.978$ & $1.96 \pm 1.0$ & $0.978$ & $1.23 \pm 0.6$ & $0.860$ & $8.3$ \\
 Ensemble Optimal* & $111 \pm 75$ & $0.979$ & $1.85 \pm 1.0$ & $0.980$ & $1.19 \pm 0.6$ & $0.871$ & $7.0$ \\ 
 Ensemble Testset & $371$ & $0.925$ & $3.02$ & $0.957$ & $2.53$ & $0.826$ & $11.5$ \\ \hline   
 DMTRL \cite{xue2018full} & $180 \pm 118$ & $0.945$ & $2.51 \pm 1.6$ & $0.925$ & $1.39 \pm 0.7$ & $0.768$ & $8.2$ \\ \hline  
\end{tabular}
\end{center}
\label{tab:results}
\end{table}

\section{Results}

All results are shown in Table~\ref{tab:results}. For Densenet121 (DN121) with 2D slice inputs, the pretrained model with segmentation regularization performs best. The difference in the median of the absolute errors of DN121 2D and DN121 2D SR is significant for areas and dimensions (Wilcoxon signed-rank test, $\alpha$ = 5\% significance level). Combining regression and classification leads to a lower performance than training separate models. For DN121 in its 3D version ($N_S = 5$), the overall performance improves slightly with respect to its 2D counterparts. Again, the difference in the median of the absolute errors of the 2D and 3D model is significant for areas and dimensions. 

Considering different architectures, improved performance can be observed for larger models. The best performing model is RX101-64d. With respect to ensembling, taking the average over all our models does not perform better than the best single model. Using our optimal subset strategy, performance is improved. When evaluating on a different test split (*), the performce difference is still large between averaging and our strategy. The difference in the median of the absolute errors for averging and our ensembling strategy is statically significant for all three indices. Our final ensembles mostly contain the models RX101-64d 2D SR, DN 2D SR variants, and DN121 3D SR. On the test set of the LVQuan19 Challenge, the performance of the ensemble is substantially lower. For reference, we include the results from DMTRL \cite{xue2018full}. Note that these results are not directly comparable, as a different number of patients and different image resolutions were used.

\section{Discussion and Conclusion}

We address LV quantification from cardiac MRI images with a focus on utilizing pretrained models. We consider a variety of deep learning models that have been successful for classification of 2D images. We adopt these models by replacing the output layer and performing direct LV indices regression from 2D MRI slices. We find a substantial increase in performance with pretrained weights, e.g., the MAE for area estimation improves from $\SI{199}{\milli\metre^2}$ to $\SI{134}{\milli\metre^2}$. This is likely tied to the new and small dataset which contains only roughly a third of the number of patients compared to most previous studies \cite{xue2018full,wang2019quantification}. At the same time, the MRI images have a higher resolution which is closer to the standard input resolution of models trained on ImageNet. Both likely lead to a substantial advantage of utilizing pretrained models.

To incorporate previous segmentation-based approaches, we propose a segmentation regularization by adding an architecture-independent decoder close to the model output. The additional segmentation loss forces the model core to learn both features for direct indices regression and myocardium segmentation. Our results indicate that including the segmentation is advantageous as we observe a statisticially significant increase in performance both for areas and dimensions. This matches insights from previous work where using both a segmentation mask and LV indices lead to improved results \cite{xu2018calculation,wang2019quantification}.

Furthermore, we address temporal dependencies by extending the existing 2D CNN models to 3D. To enable 3D CNN usage with very limited data, we use an initialization strategy where we copy pretrained 2D weights to 3D kernels. Again, we find a substantial increase in performance by relying on pretraining, see Table~\ref{tab:results}. There is a significant improvement for dimension and area regression for 3D CNNs over 2D CNNs with an MAE of $\SI{139}{\milli\metre^2}$ compared to $\SI{133}{\milli\metre^2}$ for the areas. For the cardiac phase it is notable that the 2D approach with neighboring slices performs reasonably well and only slightly worse than 3D CNNs. Overall, enabled by our initialization strategy, spatio-temporal 3D CNNs improve performance over spatial 2D CNNs for LV quantification.

Next, we consider different baseline architectures for our approach. Using larger architectures with more layers and/or more feature maps tends to improve performance over the DN121 baseline. In particular, it is notable that the highest performance increase among different models is substantially larger than the performance increase of moving from 2D to 3D. RX101-64d 2D improves the MAE by $\SI{16}{\milli\metre^2}$ over DN121 compared to a $\SI{6}{\milli\metre^2}$ decrease caused by using 3D convolutions. In the best case, one would extend the best 2D model to 3D for performance maximization, however, this is limited by GPU memory and not feasible for the larger, higher performing 2D models. Summarized, using high-performing 2D architectures can be very beneficial for LV quantification. 

Last, we combine all our models with a new ensembling technique where the best performing models were automatically selected based on cross-validation performance. The method improves performance over simply averaging predictions across all models. Also, we used separate test splits to ensure that the optimal subset selection does not implicitly overfit to the CV sets. Even for this evaluation scenario our ensembling method performs better than averaging with statistically significant performance differences. Interestingly, the selection method included both 2D and 3D CNNs, which indicates that both spatial and spatio-temporal information is important for LV indices regression. On the LVQuan19 test set, our method performs substantially worse than in our CV experiments. This indicates that the test set is very challenging and potentially differs from the training set. Similar observations were made for the last year's challenge \cite{xu2018calculation}. Thus, generalizable LV quantification remains a challenging task. Future work could incorporate our approach into other frameworks, e.g., by considering multi-task relationship learning \cite{xue2018full} or recurrent models for temporal processing.

\bibliographystyle{spmpsci}      % mathematics and physical sciences
\bibliography{egbib} 

\end{document}